\documentclass[10pt,a4paper]{article}
\usepackage{epsfig}
\begin{document}
%
%
\date{\today}
\title{Particle ratios at SPS, AGS and SIS.}
\author{Jean Cleymans$^1$, Helmut Oeschler$^2$
and Krzysztof Redlich$^{3,4}$\\[0.5cm]
$^1$Department  of  Physics,  University of Cape Town,\\
Rondebosch 7701, South Africa\\
$^2$Institut f\"ur Kernphysik, Technische Universit\"at Darmstadt, \\
D-64289 Darmstadt, Germany\\
$^3$Gesellschaft f\"ur Schwerionenforschung, D-64291 Darmstadt, Germany\\
$^4$Institute of Theoretical Physics, University of Wroc\l aw,
 Wroc\l aw, Poland\\}
\maketitle
\begin{abstract}
Ratios  of integrated particle yields 
provide the best method for 
determining  the temperature
and the chemical potential. 
The chemical freeze-out parameters obtained at  CERN/SPS,
BNL/AGS  and  GSI/SIS energies all correspond to a unique value
of  1  GeV  per  hadron  in the local rest frame of the system,
independent  of  the  beam  energy  and  of the target and beam
particles.
\end{abstract}
\section{Particle Ratios}
Ratios of integrated particle yields provide the best way to determine 
the  freeze-out values of the temperature and baryon chemical potential
of   hadronic   matter   produced  in  relativistiv  heavy  ion
collisions.
Various
effects (e.g.  flow) which can severely distort 
the momentum
spectra of the particles produced  cancel out  in 
such ratios (see e.g. \cite{jaipur}). 
The analysis of particle ratios 
is  therefore the best method to obtain reliable information on
the chemical freeze-out parameters of the hadronic final state.
Below  we  briefly  summarize  arguments supporting this
statement.
\begin{enumerate}
\item Excluded volume corrections cancel out 
in  particle  ratios.  In  most models one simply has the following
relations (see e.g. \cite{crss,greiner})
\begin{eqnarray}
\displaystyle
{N_i\over    N_j}   &=&
 {{N_i^0\over 1 + \sum_l N_l^0V_0}
   \over  {N_j^0\over 1 + \sum_l N_l^0V_0}} ,\nonumber\\
   &=&  {N_i^0\over  N_j^0},
\end{eqnarray}
where  $N_i^0$  refers  to  the distribution from a fireball at
rest which in 
the Boltzmann approximation is given by
\begin{eqnarray}
N_i^0 &=& g\int {d^3p\over (2\pi)^3} e^{-E/T}e^{\mu/T} \nonumber \\
      &=& gm_i^2TK_2(m_i/T)e^{\mu/T} .
\end{eqnarray}
Equation (1) shows that effects  due to excluded  volume corrections  
cancel  out  in particle ratios. As a word of caution : there exist 
models where these
corrections   do   not   cancel  exactly  as  above,  see  e.g.
\cite{stachel}.
\item  If the hadronic gas is made up of a 
superposition of fireballs  having different rapidities
but the same temperature and chemical potential then
the  particle  ratios  are  as if there were only one
fireball:
\begin{eqnarray}
{N_i\over    N_j}   &=&
 {\int_{-\infty}^{\infty}dy
          \int_{-Y}^YdY_{FB}~\rho(Y_{FB})
	   {dN_i^0\over dy}(y-Y_{FB})
   \over \int_{-\infty}^{\infty}dy
          \int_{-Y}^Y dY_{FB}~\rho(Y_{FB}) 
	  { dN_i^0\over dy}(y-Y_{FB})}   ,\nonumber\\
&=&{N_i^0\over N_j^0}~{\int_{-Y}^Y~dY_{FB}~\rho (Y_{FB})
   \over\int_{-Y}^Y~dY_{FB}~\rho (Y_{FB})} ,\nonumber\\
&=&  {N_i^0\over  N_j^0}.
\end{eqnarray}
Hence effects  due to a superposition of similar fireballs  cancel out.
It is of course a severe limitation that the fireballs must all
have  the  same  temperature, this has been lifted partially by
Becattini recently \cite{becattini}.
\item Transverse flow (or random walk \cite{random}) effects 
with instantaneous 
freeze-out cancel out if one considers
particle  yields  that  have  been  integrated  over transverse
momentum.
The discussion about transverse flow 
needs the following two integrals:
\begin{equation}
\int_0^\infty       dp_T
m_T^i~~K_1\left({m_T^i\over T}\cosh y_T\right)
I_0\left({ p_T\over T}\sinh y_T\right)
=m_i^2K_2({m^i\over T})\cosh y_T ,
\end{equation}
and
\begin{equation}
\int_0^\infty       dp_T
p_T~K_0\left({m_T^i\over T}\cosh y_T\right)
I_1\left({ p_T\over T}\sinh y_T\right)
=m_i^2K_2({m^i\over T})\sinh y_T   .
\end{equation}
For instantaneous freeze-out
and constant transverse velocity, one has 
\begin{eqnarray}
{N_i\over    N_j}   &=&   
 {\int_0^\infty       dp_T
m_T^i~~K_1\left({\displaystyle m_T^i\over T}\cosh y_T\right)
I_0\left({\displaystyle p_T\over T}\sinh y_T\right)
\over
\int_0^\infty       dp_T
m_T^j~~K_1\left({\displaystyle m_T^j\over T}\cosh y_T\right)
I_0\left({\displaystyle p_T\over T}\sinh y_T\right)} ,
\nonumber\\
&=&  {m_i^2~~K_2\left({\displaystyle m^i\over T}\right) 
\cosh y_T
\over
m_j^2~~K_2\left({\displaystyle m^j\over T}\right)\cosh y_T} ,
\nonumber\\
 &=&  {N_i^0\over  N_j^0}  .
\end{eqnarray}
i.e.  effects due to transverse flow cancel out if the particle
yields have been integrated over all transverse momenta.
\item For a  longitudinal expansion 
\`a  la Bjorken accompanied by a transverse
expansion, the effects 
also cancel  if one considers particle ratios integrated
over  transverse  momenta  as  shown below. The distribution 
is, in this case, given by 
\begin{eqnarray}
\left( {dN_i\over dy m_Tdm_T}\right)_{y=0}
&  &   = {g\over \pi} \int_\sigma r~dr~\tau_F(r)    \nonumber\\
& &\left\{ m_T^iI_0 \left( {p_T\sinh y_T\over T} \right)\right.
          K_1 \left( {m_T^i\cosh y_T\over T} \right) \nonumber \\
-& &\left( {\partial\tau_F\over\partial r} \right) p_T
          I_1 \left( {p_T\sinh y_T\over T} \right)
    \left. K_0 \left( {m_T^i\cosh y_T\over T} \right) \right\}
\end{eqnarray}
where  $\tau_F(r)$  is  the  proper  freeze-out time which is a
function of the radial distance, $r$.  
After integration over $m_T$   one is left with
\begin{eqnarray}
\left( {dN_i\over dy }\right)_{y=0}
= & &{g\over \pi} \int_\sigma r~dr~\tau_F(r)    \nonumber \\
& & \left\{ \cosh(y_T) - \left({\partial\tau_F\over\partial r} \right) 
\sinh (y_T) \right\} m_i^2 T K_2 \left( {m_i\over T} \right)
\end{eqnarray}
\bigskip
Since  the freeze-out parameters are the same for all particles
it follows that
\begin{equation}
{dN_i/dy\over dN_j/dy} = {N_i^0\over N_j^0} 
\end{equation}
This  relation  shows  that  in  a  model  where  the  momentum
distribution  shows  a  plateau in rapidity space and where the
flow   pattern  could  be  completely  change the  transverse
momentum distribution, the ratio of particles is still
as if one would have a single fireball at rest.
\end{enumerate}
\begin{center}
\begin{figure}
\epsfig{file=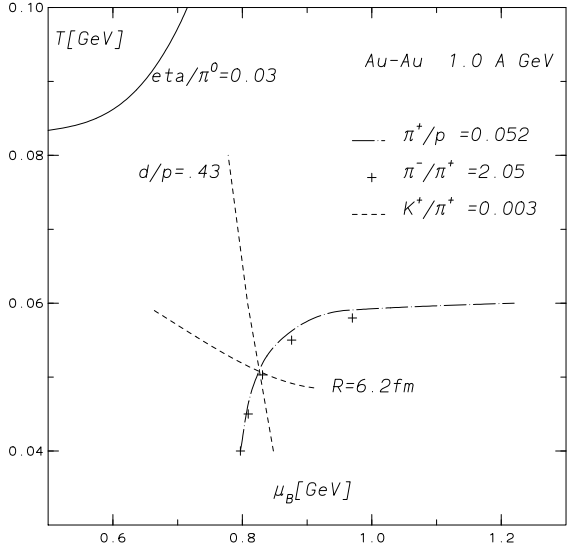,scale=0.65}
\caption{
Lines corresponding to fixed particle ratios 
in  the  $T-\mu_B$ plane. The indicated values were obtained at
the   GSI/SIS   accelerator   for   $Au-Au$   collisions  at  1
A$\cdot$GeV.}
\end{figure}
\end{center}
\section{Results}
 We show in
Fig. 1 the results obtained from an analysis \cite{oeschler} 
of $Au-Au$ collisons at an energy of 1 A$\cdot$GeV.
As  one  can  see  from an inspection of this figure there is a
consistency  with a freeze-out temperature of about 50 MeV and a
a freeze-out baryon chemical potential of about 850 MeV. A more
complete   analysis  of  the  GSI/SIS  data  can  be  found  in
\cite{oeschler}  and  in  \cite{ceks,averbeck,PBM}.  A  notable
exception  is  the  $\eta/\pi^0$  ratio  which does not fit the
expectations of the thermal model.
Combining the results from different accelerators (for a review
and references see Sollfrank \cite{sollfrank})
it can be seen that
  a  unified  description  of the hadronic
abundances  produced  in  heavy ion collisions at the CERN/SPS,
the  BNL/AGS  and  the  GSI/SIS  accelerators 
is possible \cite{prl}. This
description   covers   a   range  in  beam  energies  from  200
A$\cdot$GeV  to  below  1  A$\cdot$GeV.  As  it  turns  out the
same description can also be applied to the hadronic abundances
in  LEP  and  in  $p-p$ and $\bar{p}-p$ collisions
with slightly  different treatment of strangeness sector which
accounts for strangeness under-saturation.
 The
result
can be summarized in a surprisingly simple way: the
hadronic  composition  of  the  final state is determined by an
energy  per  hadron being approximately 1 GeV per hadron in the
rest  frame  of  the  produced  system.  This  generalizes  an
observation made a long time ago by Hagedorn \cite{hagedorn}
 for  $p-p$
collisions, namely, as one increases the
beam energy, the available energy is used to produce more
particles, but not to increase the temperature of the system.  
This  led  Hagedorn  to  the  idea  of  a  limiting
temperature.  For  heavy  ion collisions one has to take into
account  not only the temperature but also 
the finite baryon density
of  the  system,  which is described  by  the  baryon 
chemical potential
$\mu_B$.
%
This, as it was first indicated by P. Braun-Munzinger and J. Stachel 
\cite{stachel},
 leads to a freeze-out curve in the $T,\mu_B$ plane. In
Fig. 2  the  values  of the freeze-out parameters are shown, as
obtained   by  various  groups  (a summary  can  be  found  in
 \cite{sollfrank}). 
The solid line corresponds to 1 GeV
per hadron, the dashed line corresponds to 0.94 GeV per hadron.
This  energy  corresponds  to  the  chemical  freeze-out stage,
namely,  before the hadrons  decay into 
the stable hadrons.
%
Such  an  analysis,  relying  as  much  as  possible  on  fully
integrated   particle   multiplicities   was   carried   out
 for BNL/AGS and for CERN/SPS data. 
 In  Fig.   2  the  SPS  points  are indicated by open squares
 \cite{stachel,bgs}  while  the  AGS  points  are indicated by open
 circles \cite{stachel,cets,bgs}.

Data  using $Ni$ and $Au$ beams at energies between 0.8 and 1.9
A$\cdot$GeV  have  become  available  recently from the GSI/SIS
accelerator.  These  data  have attracted considerable interest
due  to  the  surprisingly  large  number of $K^-$ mesons being
produced   below  threshold.  A  very  detailed  and  extensive
discussion of these results in the framework of thermal models 
has been presented in \cite{ceks,averbeck,oeschler}. 
The results for the
freeze-out  parameters  
for  $Ni-Ni$  at  1.9  A$\cdot$GeV is shown as an open triangle
in Fig.  2.
The
points  with  the  lowest  temperature  correspond  to  $Au-Au$
collisions at 0.8 and 1.0 A$\cdot$GeV and $Ni-Ni$ collisions at
1.0 and 1.8 A$\cdot$GeV. 
and are also shown as open triangles.
\begin{center}
\begin{figure}
\epsfig{file=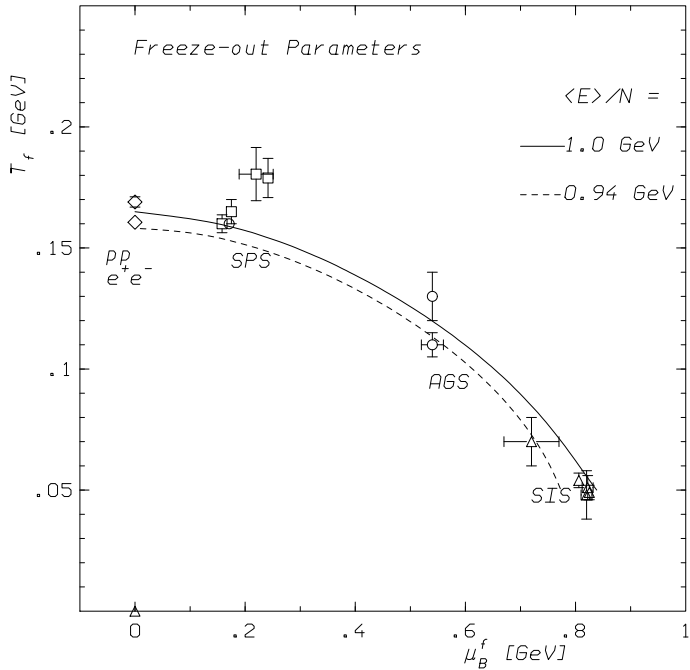,scale=0.5}
\caption{
Freeze-out   values   obtained   from  hadronic  abundances  at
CERN/SPS,  BNL/AGS  and  GSI/SIS. Also indicated are the points
obtained  from observed hadronic abundances at LEP and in $p-p$
collisions  at  CERN.  The  smooth curves correspond to a fixed
energy per hadron in the hadronic gas model (from \cite{prl}). }
\end{figure}
\end{center}
A similar   analysis   has  been  performed  in
\cite{becattini} for $e^+e^-$
annihilation into hadrons at LEP.
Since  no  baryons  are  involved here this corresponds to zero
baryon chemical potential, $\mu_B=0$. An impressive fit has
been  obtained  here  since  no less than 29 different hadronic
abundances  can  be reproduced. It is our view that such a good
agreement  cannot  simply  be  a coincidence. This analysis was
subsequently   extended \cite{becattini} to  $p-p$  and
$\bar{p}-p$  reactions  at  CERN. 
 In  this  case, 
one reproduces the Hagedorn temperature obtained many
years ago.

In the underlying hadronic gas model all these points 
can be described by a single
curve corresponding to a fixed energy per particle, $\epsilon/n$, 
which has
approximately the value of 1 GeV per particle in the hadronic gas. 
{\it This value characterizes all the final states produced by 
beams having
1  A$\cdot$GeV  all  the  way  up  to  200   A$\cdot$GeV.}
Thus, the only modification one needs to make to the concept
of Hagedorn's limiting temperature it that there exist
a "limiting" - freeze-out energy per particle of 1 GeV at which
hadrons are formed in a collision. 

This  observation  leads to a considerable
unification  in  the  description  of  the
hadronic final states produced in high 
energy collisions.
\subsection*{Acknowledgment}
\hspace*{\parindent}
We acknowledge stimulating discussions with P. Braun-Munzinger,
B. Friman, W. N\"orenberg, H. Satz  and W. Weinhold.
\end{document}